\documentclass[10pt,aps,twocolumn,prb,amsmath,amssymb,showpacs]{revtex4-1}
\usepackage{graphicx}

\pdfoutput=1

\newcommand{\eq}[1]{\begin{equation}#1\end{equation}}

\begin{document}
\title{Landau-Lifshitz theory of the longitudinal spin Seebeck effect}
\author{Silas Hoffman}
\author{Koji Sato}
\author{Yaroslav Tserkovnyak}
\affiliation{Department of Physics and Astronomy, University of California, Los Angeles, California 90095, USA}

\begin{abstract}
Thermal-bias-induced spin angular momentum transfer between a paramagnetic metal and ferromagnetic insulator is studied theoretically based on the stochastic Landau-Lifshitz-Gilbert (LLG) phenomenology. Magnons in the ferromagnet establish a nonequilibrium steady state by equilibrating with phonons via bulk Gilbert damping and electrons in the paramagnet via spin pumping, according to the fluctuation-dissipation theorem. Subthermal magnons and the associated spin currents are treated classically, while the appropriate quantum crossover is imposed on high-frequency magnetic fluctuations. We identify several length scales in the ferromagnet, which govern qualitative changes in the dependence of the thermally-induced spin current on the magnetic film thickness.
\end{abstract}



\maketitle

\section{Introduction}

Over the past three decades, spintronics has evolved from a focus on equilibrium phenomena in magnetic heterostructures, such as giant magnetoresistance \cite{baibichPRL88} and interlayer exchange interactions,\cite{binaschPRB89,*parkinPRL90} to dynamic processes, such as spin-transfer torque \cite{slonczewskiJMMM96,*bergerPRB96,tsoiPRL98,*myersSCI99} and spin pumping,\cite{mizukamiJJAP01,*urbanPRL01,*heinrichPRL03,tserkovPRL02sp,*tserkovRMP05}  and, more recently, nonequilibrium thermodynamics, heralded by the spin Seebeck effect \cite{uchidaNAT08,*uchidaNATM10,*jaworskiNATP10,xiaoPRB10,adachiRPP13} and thermally-induced motion of domain walls. \cite{kovalevPRB09te,*bauerPRB10,*hinzkePRL11,*yanPRL11,*kovalevEPL12,torrejonPRL12,*jiangPRL13dw} From a practical standpoint, magnetic nanostructures are useful for field sensing and nonvolatile information storage,\cite{wolfSCI01} where  magnetoresistance is paramount for the readout, while current-induced spin torques are useful for fast and scalable bit switching.\cite{akermanSCI05,*maekawaBOOK12}

One rapidly-developing  avenue of research concerns out-of-equilibrium spin phenomena in \textit{insulating} systems, where spin is carried by collective excitations, such as spin waves (magnons), rather than electronic quasiparticles. To this end, spin waves in the ferrimagnetic insulator yttrium iron garnet (YIG) appear particularly promising as they suffer from a remarkably low Gilbert damping (at microwave frequencies), $\alpha\sim10^{-4}$, and the host material has Curie temperature of $\sim500$~K, thus remaining magnetic at room temperature.\cite{bhagatPSS73} Spin waves in YIG have recently been shown to undergo room-temperature Bose-Einstein condensation under nonlinear microwave pumping,\cite{demokritovNAT06} exhibit large spin pumping into adjacent conductors,\cite{sandwegAPL10,*sandwegPRL11,heinrichPRL11,burrowesAPL12} manifest the longitudinal spin Seebeck effect,\cite{uchidaAPL10} and efficiently move domain walls under small thermal gradients.\cite{jiangPRL13dw} These phenomena hold promise for integrated circuits based on nonvolatile magnetic elements\cite{allwoodSCI05,*ikedaIEEEE07,*khitunIEEEM08} with essentially no Ohmic losses and thus very low dissipation.

Furthermore, thermal control of magnetic dynamics and spin currents\cite{bauerNATM12} provides an attractive alternative to voltage control, especially since magnons, which are neutral objects, can respond more directly to temperature gradients. The spin Seebeck effect, i.e., the generation of thermal spin current between magnetic insulators and normal metals, is the basic phenomenon of central interest in this context. The purpose of this paper is to develop a systematic semi-phenomenological approach to this problem, based on the Landau-Lifshitz-Gilbert (LLG) theory of ferromagnetic dynamics,\cite{landauBOOKv9,*gilbertIEEEM04} departing from the spin-pumping\cite{tserkovPRL02sp} perspective on the interaction between electrons and magnons at ferromagnetic-insulator$\mid$normal-metal interfaces put forward in Ref.~\onlinecite{xiaoPRB10}. In Sec.~\ref{DS}, we comment on how the theory could be expanded to account for magnon and phonon kinetics when the standard LLG phenomenology fails.

\section{Ferromagnetic bulk dynamics}

In the ferromagnetic bulk, away from the Curie temperature, magnetic dynamics are described by the stochastic  LLG equation\cite{landauBOOKv9}
\begin{equation}
\partial_t\mathbf{m}=-\gamma\mathbf{m}\times(\mathbf{H}_{\rm eff}+\mathbf{h}_l)+\alpha\mathbf{m}\times\partial_t\mathbf{m}\,,
\end{equation}
where $\mathbf{m}=\mathbf{M}/M_s$ is the unit-vector magnetization direction ($M_s=|\mathbf{M}|$ being the saturated magnetization magnitude), $\gamma$ (minus) the gyromagnetic ratio ($\gamma>0$ for free electrons), $\alpha$ dimensionless Gilbert damping constant,
\begin{equation}
\mathbf{H}_{\rm eff}\equiv-\delta_\mathbf{M}F=H_a\mathbf{z}+A_x\nabla^2\mathbf{m}+\mathbf{H}_r
\label{H}
\end{equation}
the effective field (consisting of applied field $H_a$ in the $z$ direction, exchange field $\propto A_x$, and relativistic corrections $\mathbf{H}_r$ that include dipolar interactions and crystalline anisotropies), and $\mathbf{h}_l$ random Langevin field with correlator\cite{brownPR63}
\begin{equation}
\left\langle h_{l,i}(\mathbf{r},t)h_{l,j}(\mathbf{r}',t')\right\rangle=\frac{2\alpha}{\gamma M_s}k_BT(\mathbf{r})\delta_{ij}\delta(\mathbf{r}-\mathbf{r}')\delta(t-t')\,,
\label{hh}
\end{equation}
in accordance with the fluctuation-dissipation theorem. We are interested at intermediate temperatures: much lower than the Curie temperature, such that the Landau-Lifshitz phenomenology based on the directional magnetization dynamics [SO(3) nonlinear $\sigma$ model] is appropriate, while not too low such that the classical theory can be used as a starting point. We will, furthermore, neglect $\mathbf{H}_r$ in Eq.~\eqref{H}, for simplicity, which is justified when $k_BT\gg\hbar\gamma M_s$. The Langevin correlator \eqref{hh} is white at frequencies $\omega\ll k_BT/\hbar$, corresponding to classical behavior. In Sec.~\ref{QC}, we will adapt our theory to account for quantum fluctuations at $\omega\gtrsim k_BT/\hbar$, by matching with the fully quantum treatment of Ref.~\onlinecite{benderPRL12}.

In order to streamline discussion of the spin transfer, let us switch from the magnetization to the spin density:
\begin{equation}
\mathbf{s}\equiv s\mathbf{n}=-\frac{M_s}{\gamma}\mathbf{m}\,,
\label{nm}
\end{equation}
where $s=M_s/\gamma$ is the saturated spin density and $\mathbf{n}=-\mathbf{m}$ its direction. The LLG equation then becomes
\begin{equation}
s(1+\alpha\mathbf{n}\times)\partial_t\mathbf{n}+\mathbf{n}\times(H\mathbf{z}+\mathbf{h})+\partial_i\mathbf{j}_{s,i}=0\,,
\label{LLG}
\end{equation}
where
\begin{equation}
\mathbf{j}_{s,i}=-A\mathbf{n}\times\partial_i\mathbf{n}
\label{js}
\end{equation}
is identified as the magnetic spin current and
\begin{equation}
\left\langle h_i(\mathbf{r},t)h_j(\mathbf{r}',t')\right\rangle=2\alpha sk_BT(\mathbf{r})\delta_{ij}\delta(\mathbf{r}-\mathbf{r}')\delta(t-t')\,,
\label{hhs}
\end{equation}
where $i$ and $j$ stand for the Cartesian coordinates. Here, $H\equiv M_sH_a$ and $A\equiv M_sA_x$. In equilibrium, $\mathbf{n}=-\mathbf{z}$, assuming the applied field $H_a>0$.

\begin{figure}[pt]
\includegraphics[width=0.6\linewidth]{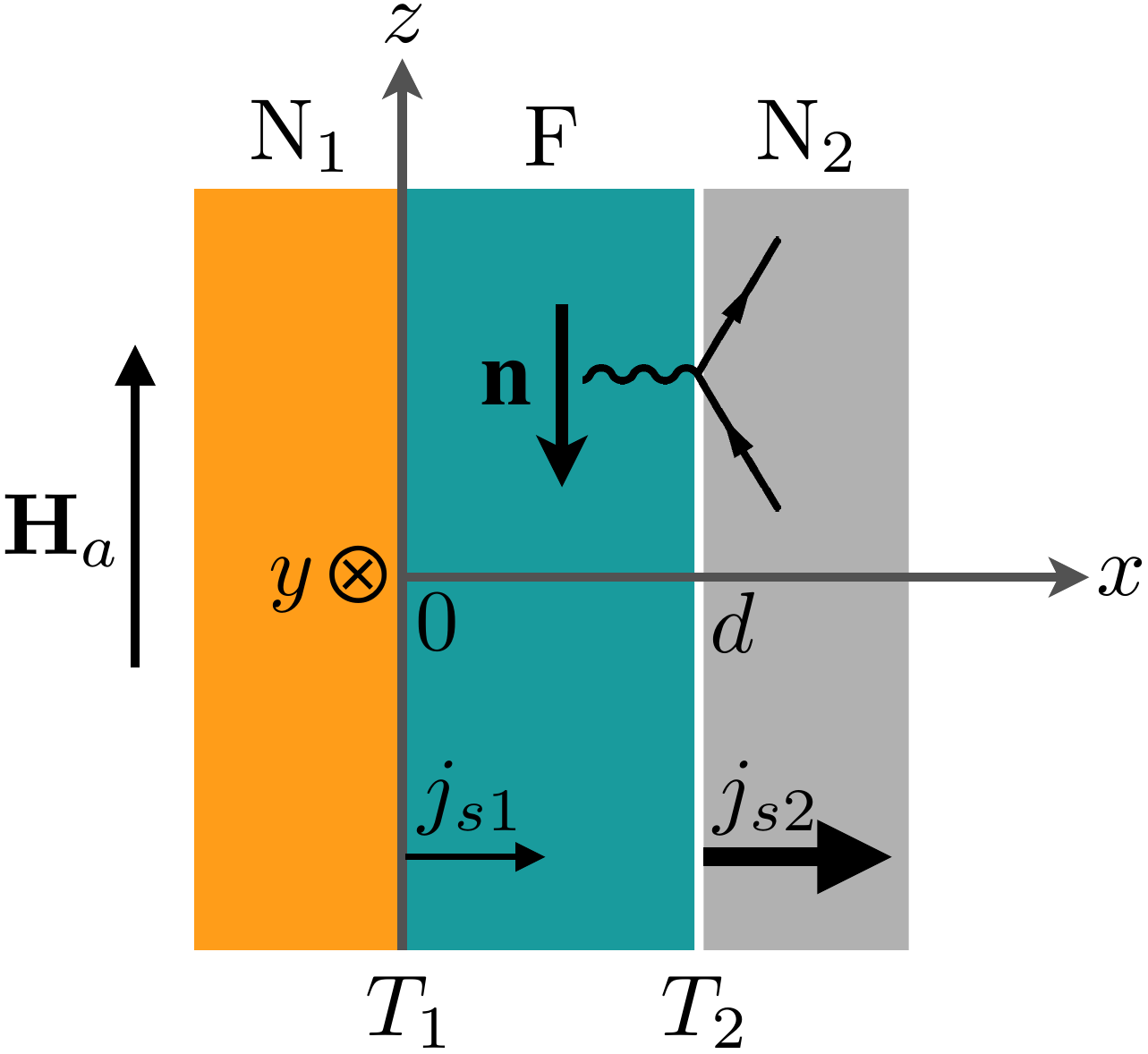}
\caption{Schematic of an {N$_1$}$\mid$F$\mid${N$_2$} sandwich structure studied in this paper. The normal-metal layer N$_1$ is treated as a poor spin sink, which blocks spin current, $j_{s1}\approx0$. The normal-metal layer N$_2$, on the other hand, is a perfect spin sink, thus establishing a thermal contact between its itinerant electrons and magnons in the ferromagnetic insulator (F), which results in spin current $j_{s2}\equiv j_s$. We assume the phonons in the F layer follow a linear temperature profile from $T_1$ at $x=0$ ({N$_1$}$\mid$F interface) to $T_2$ at $x=d$ (F$\mid${N$_2$} interface), corresponding to electron temperatures in N$_1$ and N$_2$, respectively.}
\label{sc}
\end{figure}

In this paper, we focus on the trilayer heterostructure depicted in Fig.~\ref{sc}. The temperature $T(\mathbf{r})$ entering Eq.~\eqref{hh} is taken to correspond to the $x$-dependent phonon temperature inside of the ferromagnetic film,
\begin{equation}
T(x)=T_1+\frac{x}{d}(T_2-T_1)\,,
\label{Tx}
\end{equation}
assuming Gilbert damping stems from the local magnon-phonon scattering. (We will revisit this assumption in Sec.~\ref{DS}.)

\section{Boundary conditions}

The boundary conditions for a ferromagnet sandwiched between two normal metals need to be similarly constructed to account both for deterministic\cite{tserkovPRL02sp,*tserkovRMP05} and stochastic \cite{forosPRL05} spin-transfer torques. We will start with the former and then include the latter according to the fluctuation-dissipation theorem.

We assume the spin current is blocked by the N$_1$ layer at $x=0$, due to its weak spin-relaxation rate:
\begin{equation}
\mathbf{j}_{s,x}=0~~~(x=0)\,.
\label{j0}
\end{equation}
In other words, the spin current pumped across the F$\mid${N$_1$} interface is balanced by an equal backflow.\cite{tserkovPRL02sp} For our purposes, N$_1$ can thus be replaced by an insulator, as long as it makes a good thermal contact with phonons in the ferromagnet. A net spin current across the F$\mid$N$_2$ interface, on the other hand, is allowed, if we treat N$_2$ as a perfect spin sink:\cite{tserkovPRL02sp}
\begin{equation}
\mathbf{j}_{s,x}=\frac{\hbar g^{\uparrow\downarrow}}{4\pi}\mathbf{n}\times\frac{d\mathbf{n}}{dt}~~~(x=d)\,,
\label{jd}
\end{equation}
where $g^{\uparrow\downarrow}$ is the real part of the dimensionless interfacial spin-mixing conductance (per unit area).\cite{brataasPRL00} We disregard the imaginary part of the spin-mixing conductance, since it governs the typically smaller\cite{tserkovPRL02sp} nondissipative spin-current component $\propto d\mathbf{n}/dt$, which vanishes over a cycle of precession. A specific realization for such an {N$_1$}$\mid$F$\mid${N$_2$} trilayer could be provided by the Cu$\mid$YIG$\mid$Pt combination, where Cu (Pt) is a light (heavy) element with weak (strong) spin-orbit interaction. (We will generalize our findings to arbitrary {N$_1$}$\mid$F$\mid${N$_2$} trilayers, such as symmetric Pt$\mid$YIG$\mid$Pt type structures or general asymmetric structures, in Sec.~\ref{Sum}.)

For the two {N$_1$}$\mid$F$\mid${N$_2$} interfaces, we correspondingly have the following (deterministic) boundary conditions for magnetic dynamics (as $T\to0$):
\begin{equation}
\left\{
\begin{array}{cc}\displaystyle\partial_x\mathbf{n}=0\,, & x=0\\\displaystyle A\partial_x\mathbf{n}+\frac{\hbar g^{\uparrow\downarrow}}{4\pi}\partial_t\mathbf{n}=0\,, & x=d\end{array}\right.\,,
\end{equation}
reflecting continuity of spin current, which is given by Eq.~\eqref{js} inside the ferromagnet and Eqs.~\eqref{j0} and \eqref{jd} in N$_1$ and N$_2$, respectively, across the corresponding interfaces. Since spin pumping \eqref{jd} affects magnetic dynamics similarly to Gilbert damping,\cite{tserkovPRL02sp} it is accompanied with a similar stochastic term.\cite{forosPRL05} The latter can be accounted for by modifying the boundary condition at $x=d$:
\begin{equation}
A\partial_x\mathbf{n}+\frac{\hbar g^{\uparrow\downarrow}}{4\pi}\partial_t\mathbf{n}+\mathbf{h}'=0\,,
\label{BC}
\end{equation}
where
\begin{equation}
\left\langle h'_i(\boldsymbol{\rho},t)h'_j(\boldsymbol{\rho}',t')\right\rangle=\frac{\hbar g^{\uparrow\downarrow}}{2\pi}k_BT_2\delta_{ij}\delta(\boldsymbol{\rho}-\boldsymbol{\rho}')\delta(t-t')\,
\label{hhp}
\end{equation}
and $\boldsymbol{\rho}=(y,z)$ is the two-dimensional position along the interface at $x=d$. The Langevin correlator strength is proportional to the electron temperature $T_2$ at the F$\mid$N$_2$ interface, since the noise originates in the thermal fluctuations of electronic spin currents in N$_2$.

The spin Seebeck effect is embodied in the thermal-averaged spin current flowing through the F$\mid$N$_2$ interface:\cite{xiaoPRB10}
\begin{equation}
\mathbf{j}_{s,x}=-A\mathbf{n}\times\partial_x\mathbf{n}=\mathbf{n}\times\left(\frac{\hbar g^{\uparrow\downarrow}}{4\pi}\partial_t\mathbf{n}+\mathbf{h}'\right)\,.
\end{equation}
Since our system is axially symmetric with respect to the $z$ axis, it is convenient to switch to complex notation: $n\equiv n_x-in_y$. Thermal spin-current density, $\langle\mathbf{j}_{s,x}\rangle=j_s\mathbf{z}$, can thus be written for small-angle dynamics (relevant at temperatures well below the Curie temperature) as
\begin{equation}
j_s=A\,{\rm Im}\left.\left\langle n^\ast\partial_xn\right\rangle\right|_{x=d}\,.
\end{equation}
Exploiting, furthermore, translational invariance in the $yz$ plane, we find in the steady state:
\begin{equation}
j_s=A\,{\rm Im}\int\frac{d^2\mathbf{q}d\omega}{(2\pi)^3}\frac{\left\langle n(\mathbf{q},\omega)^\ast\partial_xn(\mathbf{q}',\omega')\right\rangle}{(2\pi)^3\delta(\mathbf{q}-\mathbf{q}')\delta(\omega-\omega')}\,,
\label{jnn}
\end{equation}
where
\begin{equation}
n(\mathbf{q},\omega)=\int d^2\boldsymbol{\rho}dte^{i(\omega t-\mathbf{q}\cdot\boldsymbol{\rho})}n(\boldsymbol{\rho},d,t)
\end{equation}
is the Fourier transform over $\boldsymbol{\rho}$ and time $t$. The delta functions in the denominator of Eq.~\eqref{jnn} cancel delta functions that factor out of the numerator when evaluating the average $\langle\dots\rangle$ (with the remaining integrand independent of $\mathbf{q}'$ and $\omega'$). Similarly transforming Langevin correlators, Eqs.~\eqref{hhs} and \eqref{hhp}, we have:
\begin{align}
\left\langle h(x,\mathbf{q},\omega)^\ast h(x',\mathbf{q}',\omega')\right\rangle=&4(2\pi)^3\alpha sk_BT(x)\nonumber\\
&\hspace{-1.5cm}\times\delta(x-x')\delta(\mathbf{q}-\mathbf{q}')\delta(\omega-\omega')\,,
\end{align}
for the bulk and
\begin{equation}
\left\langle h'(\mathbf{q},\omega)^\ast h'(\mathbf{q}',\omega')\right\rangle=4(2\pi)^3\alpha'sk_BT_2\delta(\mathbf{q}-\mathbf{q}')\delta(\omega-\omega')
\end{equation}
for the F$\mid$N$_2$ interface, defining
\begin{equation}
\alpha'\equiv\frac{\hbar g^{\uparrow\downarrow}}{4\pi s}\,,
\label{ap}
\end{equation}
which has dimensions of length. $\alpha'/d$ is the enhanced Gilbert damping for a monodomain precession of the ferromagnetic film.\cite{tserkovPRL02sp}

\section{Spin Seebeck coefficient}

We now have all the necessary ingredients in order to evaluate the (longitudinal) spin Seebeck coefficient (which has units of inverse length squared)\cite{Note1}
\begin{equation}
S\equiv\frac{j_s}{k_B(T_1-T_2)}
\label{S}
\end{equation}
of the {N$_1$}$\mid$F$\mid${N$_2$} structure shown in Fig.~\ref{sc}. To simplify our subsequent analysis, let us optimize the notation, as follows. The stochastic LLG equation \eqref{LLG} in the film bulk is written as
\begin{equation}
A(\partial_x^2-\kappa^2)n(x,\mathbf{q},\omega)=h(x,\mathbf{q},\omega)\,,
\label{LLGc}
\end{equation}
after linearizing transverse dynamics and Fourier transforming it in the $yz$ plane and time. Here,
\begin{equation}
\kappa^2\equiv q^2+\frac{H-(1+i\alpha)s\omega}{A}\,.
\label{BCc}
\end{equation}
The stochastic boundary condition at $x=d$, Eq.~\eqref{BC}, in this notation is
\begin{equation}
A(\partial_x-\kappa')n(x,\mathbf{q},\omega)=-h'(\mathbf{q},\omega)~~~(x=d)\,,
\end{equation}
while $\partial_xn=0$ at $x=0$. Here,
\begin{equation}
\kappa'\equiv i\frac{\alpha's\omega}{A}\,.
\label{kp}
\end{equation}

Eqs.~\eqref{LLGc}-\eqref{kp} now form a closed system of inhomogeneous linear differential equations, with source terms given by stochastic fields $h$ and $h'$. These are straightforward to solve for $n$ using Green's functions. Substituting the solution for $n$ into Eq.~\eqref{jnn}, we find, after some algebra, the spin Seebeck coefficient \eqref{S}:
\begin{align}
S=&\frac{\alpha\alpha's^2}{2\pi^3A^2d}\int_{-\infty}^\infty d^2\mathbf{q}\int_{-\infty}^\infty d\omega \omega\nonumber\\
&\hspace{1cm}\times\int_0^ddx x\left|\frac{\cosh[\kappa(x-d)]}{\kappa\sinh(\kappa d)-\kappa'\cosh(\kappa d)}\right|^2\,.
\end{align}
Integrating over the longitudinal coordinate $x$, this finally becomes
\begin{align}
\label{SS}
S=\frac{\alpha\alpha's^2}{8\pi^3A^2d}&\int^\infty_{-\infty} d^2\textbf{q}\int^\infty_{-\infty}\frac{d\omega\omega}{\left|\kappa\sinh(\kappa d)-\kappa'\cosh(\kappa d)\right|^2}\nonumber\\
&\times\left[\frac{\sin^2(\kappa_i d)}{\kappa_i^2}+\frac{\sinh^2(\kappa_r d)}{\kappa_r^2}\right]\,,
\end{align}
where $\kappa_r$ ($\kappa_i$) is the real (imaginary) part of $\kappa$. Eq.~\eqref{SS} is our central results and the main departure point for the subsequent analysis.

Let us, for convenience, define the following length scales:
\begin{equation}
\xi\equiv\sqrt{\frac{A}{H}}
\end{equation}
is the \textit{magnetic exchange length,}
\begin{equation}
l'\equiv\frac{\alpha'}{\alpha}
\end{equation}
is the \textit{spin-pumping length} (i.e., the F thickness at which the monodomain Gilbert damping enhancement due to spin pumping\cite{tserkovPRL02sp} equals the intrinsic damping),
\begin{equation}
\lambda\equiv\sqrt{\frac{\hbar A}{sk_BT}}
\end{equation}
is the \textit{thermal de Broglie wavelength} (in the absence of applied field), where $T$ is the ambient temperature, and
\begin{equation}
l\equiv\frac{\lambda}{\alpha}
\end{equation}
is the \textit{decay length} for thermal magnons in the bulk. In this notation,
\begin{equation}
\kappa^2=q^2+\frac{1}{\xi^2}-\frac{1+i\alpha}{\lambda^2}\frac{\hbar\omega}{k_BT}\,.
\end{equation}

\section{Quasiparticle Approximation}

In the following, we are primarily interested in the thickness, $d$, dependence of the spin Seebeck coefficient, $S$, assuming the the length-scale hierarchy $\lambda\ll l'\ll l$. In YIG, for example, taking\cite{bhagatPSS73} $4\pi M_s\sim2$~kG, $A\sim1/2\times10^{-6}$~erg/cm, and $\alpha\sim10^{-4}$, we find the following lengths: (1) $\lambda\lesssim1$~nm, at room temperature, (2) $l'\sim100$~nm, taking $g^{\uparrow\downarrow}\sim10^{14}$~cm$^{-2}$ from Ref.~\onlinecite{heinrichPRL11} and proportionately larger $l'$ with $g^{\uparrow\downarrow}\sim5\times10^{14}$~cm$^{-2}$ from Ref.~\onlinecite{burrowesAPL12} ($g^{\uparrow\downarrow}$ is very sensitive to the preparation and quality of the YIG$\mid$metal interfaces), (3) $l\lesssim10$~$\mu$m, and (4) $\xi\sim10$~nm at $1$~kG (corresponding to typical magnetostatic fields).

\begin{figure}[pt]
\includegraphics[width=0.85\linewidth,clip=]{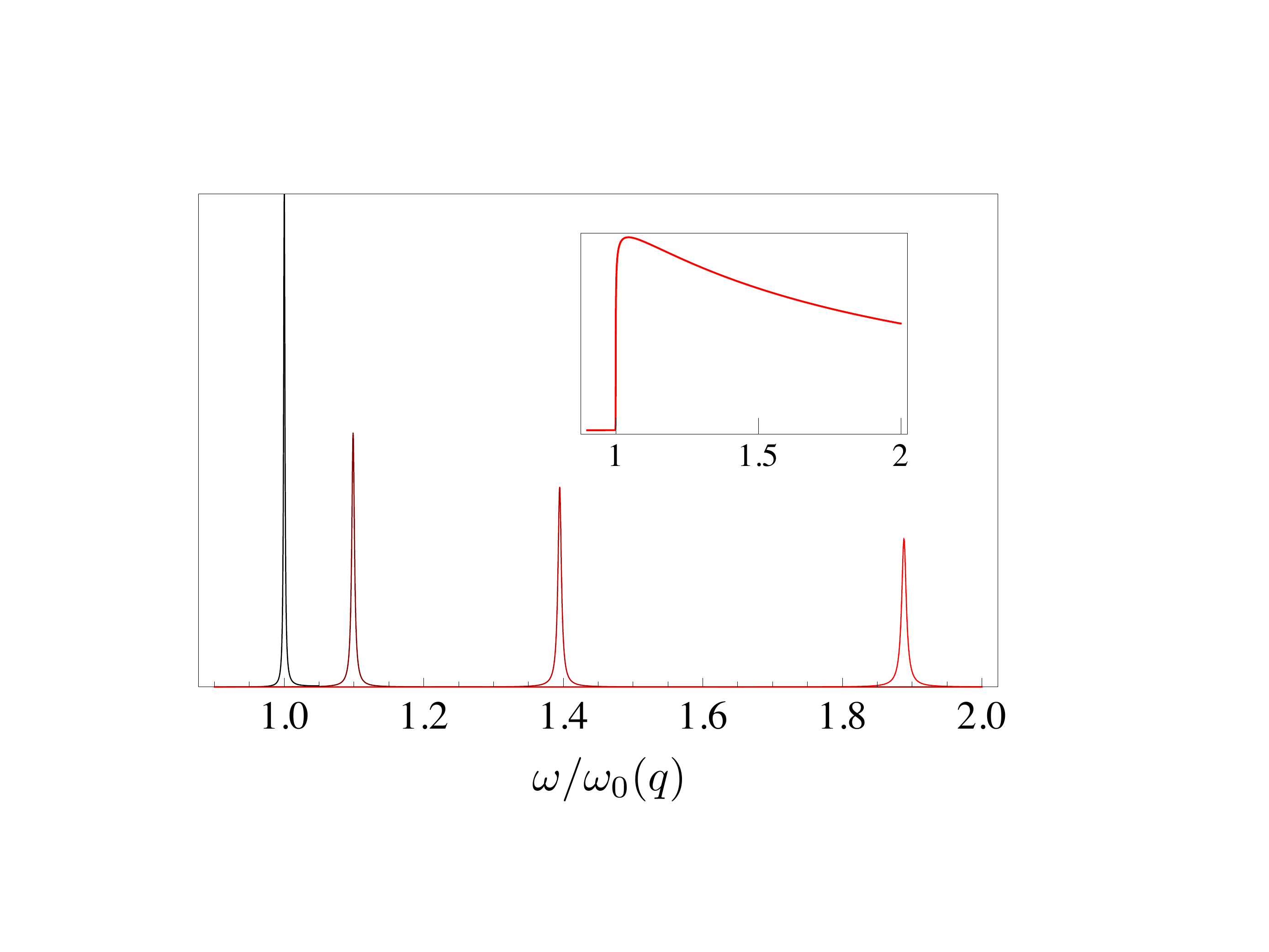}
\caption{Integrand of the spin Seebeck coefficient, Eq.~\eqref{SS}, in arbitrary units, illustrating the first four quasiparticle resonances, $n=0,1,2,3$, according to Eq.~\eqref{w}, for a fixed $q$. We set $\alpha=10^{-4}$, $\alpha'=10^{-2}$~nm, $q^2+\xi^{-2}=1$~nm$^{-2}$, and $d=10$~nm in the main plot. The inset shows the essentially continuum spectrum when $d=100$~$\mu$m, keeping other parameters unmodified.}
\label{magnon_peaks}
\end{figure}

We start by performing integration over frequency $\omega$ in Eq.~\eqref{SS} in the limit of low damping (both intrinsic and spin pumping). In this case, the integrand is peaked at $\sinh(\kappa d)\approx0$, corresponding to
\eq{\omega_n(q)=\frac{A}{s}\left(q^2+\frac{n^2\pi^2}{d^2}+\frac{1}{\xi^2}\right)\,,\label{w}}
where $n=0,1,2,3,\dots$ labels magnon subbands [not to be confused with unit vector $\mathbf{n}$ introduced in Eq.~\eqref{nm}]. These resonances are well separated when their width is much smaller than their spacing, allowing for a quasiparticle treatment of the energy integral. For the bulk damping, this condition is $\alpha \ll 1/d\sqrt{1/\lambda^2-1/\xi^2}$, for thermal magnons. Additionally, the occupation of magnons is exponentially suppressed when the temperature is smaller than the gap, i.e., $k_B T/\hbar\lesssim\omega_0\equiv H/s$. Therefore, we are  interested in the opposite regime, when the thermal de Broglie wavelength is smaller than the magnetic exchange length, $\lambda<\xi$. Thus in the regime $d\ll l$, these resonances are well-defined quasiparticle peaks corresponding to monodomain precession ($n=0$) and standing waves ($n>0$) along the longitudinal direction. See Fig.~\ref{magnon_peaks}. This allows us to evaluate the spin Seebeck coefficient  by summing the contributions from individual magnon modes, when $d\ll l$.

Expanding around $\omega_0$, the contribution from the lowest energy resonance is
\begin{equation}
S_0=\frac{\alpha\alpha'}{4\pi^3d}\int^\infty_{-\infty} d^2\textbf{q}\int^\infty_{-\infty}\frac{d\omega\omega}{(\omega-\omega_0)^2+\left(\alpha+\alpha'/d\right)^2\omega^2}\,,
\end{equation}
which can be readily integrated over frequency:
\begin{align}
S_0=&\frac{\alpha\alpha'}{4\pi^2d}\int^\infty_{-\infty} \frac{d^2\textbf{q}}{(\alpha+\alpha'/d)\left[1+(\alpha+\alpha'/d)^2\right]}\nonumber\\
\approx&\frac{\alpha'/d}{1+l'/d}\int^\infty_{-\infty} \frac{d^2\textbf{q}}{(2\pi)^2}\,,
\label{S0}
\end{align}
where we have assumed small damping, $\alpha+\alpha'/d\ll1$. Similarly, we find for the $n>0$ subbands:\cite{Note2}
\eq{S_n\approx\frac{2\alpha'/d}{1+2l'/d}\int^\infty_{-\infty}\frac{d^2\textbf{q}}{(2\pi)^2}\,.
\label{Sn}}
The total Seebeck coefficient in the quasiparticle approximation is thus
\begin{equation}
S=S_0+\sum_{n>0} S_n\,.
\label{SDgL}
\end{equation}
The wave vector $\mathbf{q}$ in integrals \eqref{S0} and \eqref{Sn} should be bounded by requiring that $\hbar\omega_n(q)<k_BT$, at which point our classical treatment breaks down, as discussed in the next section. The spurious ultraviolet divergence is eliminated by cutting off at a total three-dimensional wave number, $q_c=\sqrt{1/\lambda^2-1/\xi^2}$, corresponding to energy $k_B T$. While a more careful quantum treatment for large wave numbers is constructed in the next section, we expect a crude cutoff to adequately capture the behavior of the Seebeck coefficient, up to an overall factor of order one.

Allow us to momentarily focus on the regime when $d\gg\lambda$, where the quasiparticle peaks are dense (recovering three-dimensional behavior), so that
\begin{equation}
S\approx\sum_{n>0}S_n\approx\frac{2\alpha'}{1+2l'/d}\int_{q_c}\frac{d^3\textbf{q}}{(2\pi)^3}=\frac{\alpha'}{1+2l'/d}\frac{q_c^3}{3\pi^2}
\label{SSS}
\end{equation}
and, therefore,
\eq{
S=\left(\frac{1}{\lambda^2}-\frac{1}{\xi^2}\right)^{3/2}\times\left\{ \begin{array}{cc}
\displaystyle\frac{\alpha d}{6\pi^2}\,,&~~~d\ll l'\vspace{2mm}\\

\displaystyle\frac{\alpha'}{3\pi^2}\,,&~~~d\gg l'
\end{array} \right..
}
That is, when the thickness of the ferromagnet is much smaller or larger than spin-pumping length, the spin current scales linearly with $d$ or is $d$ independent, respectively.

If the quasiparticle peaks are not dense, $d\sim\lambda$, finite-size effects are important, reflecting individual magnon subbands. In the extreme low-temperature case when $d\ll\lambda$, only monodomain precession along the longitudinal direction contributes to the Seebeck coefficient: $S\approx S_0$. If the transverse dimensions are also much smaller than $\lambda$, the full volume of the (nano)magnet undergoes stochastic monodomain precession, and
\eq{
S=\frac{\alpha'}{V}\frac{1}{1+l'/d}\,,
\label{Sm}
}
where we have retained only one mode associated with the transverse momentum in Eq.~\eqref{S0}. $V$ here is the volume of the F layer. This coincides with the spin Seebeck coefficient for a monodomain obtained in Ref.~\onlinecite{xiaoPRB10} [defining $(T_1+T_2)/2\to T_F$, $T_2\to T_N$, and $\alpha'/d\rightarrow\alpha'$, to match their notation].

Finally, for largest thicknesses $d\gg l$, the quasiparticles are no longer well-defined (see inset of Fig.~\ref{magnon_peaks}) and the above analysis cannot be applied. Because the thickness is beyond the magnon propagation length, only magnons within a distance $l$ from the F$\mid$N$_2$ interface contribute to the spin current, which should, therefore, be independent of thickness, $d$, for a fixed thermal gradient, $(T_1-T_2)/d$. Since, in this regime, the magnon propagation length is the largest length scale in the problem, we can send $d\rightarrow\infty$ in Eq.~\eqref{SS}, which gives
\eq{
S=\frac{\alpha\alpha' s^2}{8\pi^3A^2d}\int^{\infty}_{-\infty} d^2\textbf{q} \int^{\infty}_{-\infty}\frac{d\omega\omega}{\kappa_r^2\left|\kappa-\kappa'\right|^2}\,.
\label{Sdinfty}
}
The integrand, which can be evaluated numerically, is independent of thickness, and, therefore, $S\propto1/d$, as expected. 

\section{Quantum crossover}
\label{QC}

Our classical Langevin theory needs to be appropriately modified when approaching magnon frequencies of $\hbar\omega\sim k_BT$. On the one hand, this is an important limit, as the spin transport is dominated by thermal magnons in our model. On the other hand, the classical theory is inadequate for the treatment of quantum fluctuations that dominate at high (on the scale of the ambient temperature) frequencies.

To this end, we use a quantum-mechanical result\cite{benderPRL12} for the thermal spin current, which is exact for a tunneling spin-exchange Hamiltonian at an F$\mid$N interface:\cite{Note3}
\begin{equation}
j_s=-4k_B\delta T\alpha'\int_{\epsilon_0}^\infty d\epsilon\epsilon D(\epsilon)\beta^2\frac{\partial n_{\rm BE}(\beta\epsilon)}{\partial\beta}\,,
\label{QS}
\end{equation}
where $D(\epsilon)$ is the magnon density of states, $n_{\rm BE}(x)\equiv(e^{x}-1)^{-1}$ is the Bose-Einstein distribution, $\beta\equiv1/k_BT$, and $\delta T$ is the temperature drop across the F$\mid$N interface (assuming magnons are equilibrated to a uniform temperature $T$, such that $\lambda\ll d$). This limit can be directly compared to our Eqs.~\eqref{S} and \eqref{SS}, by first sending $\alpha'\to0$ in the integrand (thus reproducing the weak F$\mid$N contact and allowing for the magnons in F to equilibrate with phonons) and then sending $\alpha\to0$ [such that the magnon spectral properties are unaffected by Gilbert damping, as assumed in the derivation of Eq.~\eqref{QS}]. The magnons are correspondingly equilibrated to the average phonon temperature $(T_1+T_2)/2$, such that we identify $\delta T=(T_1-T_2)/2$, in the present notation. According to Eq.~\eqref{SSS}, our semiclassical spin current becomes in this limit
\eq{
j_s\to2k_B(T_1-T_2)\alpha'\int\frac{d^3\mathbf{q}}{(2\pi)^3}\,.
}
This agrees with Eq.~\eqref{QS} in the limit $\epsilon\ll k_BT$\,, where
\eq{
-\int^\infty_{\epsilon_0}d\epsilon\epsilon D(\epsilon)\beta^2\frac{\partial n_{\rm BE}(\beta\epsilon)}{\partial\beta}\to\int^\infty_{\epsilon_0} d\epsilon D(\epsilon)\equiv\int\frac{d^3\mathbf{q}}{(2\pi)^3}\,.
}

We conclude that the classical-to-quantum crossover can be accounted for by inserting the factor
\begin{equation}
\label{QF}
-\epsilon\beta^2\frac{\partial n_{\rm BE}(\beta\epsilon)}{\partial\beta}=\left[\frac{\beta\hbar\omega/2}{\sinh(\beta\hbar\omega/2)}\right]^2\equiv F(\beta\hbar\omega)
\end{equation}
in the energy integrand of Eq.~\eqref{SS}, which effectively cuts off the contribution from magnons with energy $\epsilon\equiv\hbar\omega\gg k_BT$.

\section{Results}
\label{Sum}

\begin{figure}[pt]
\includegraphics[width=\linewidth,clip=]{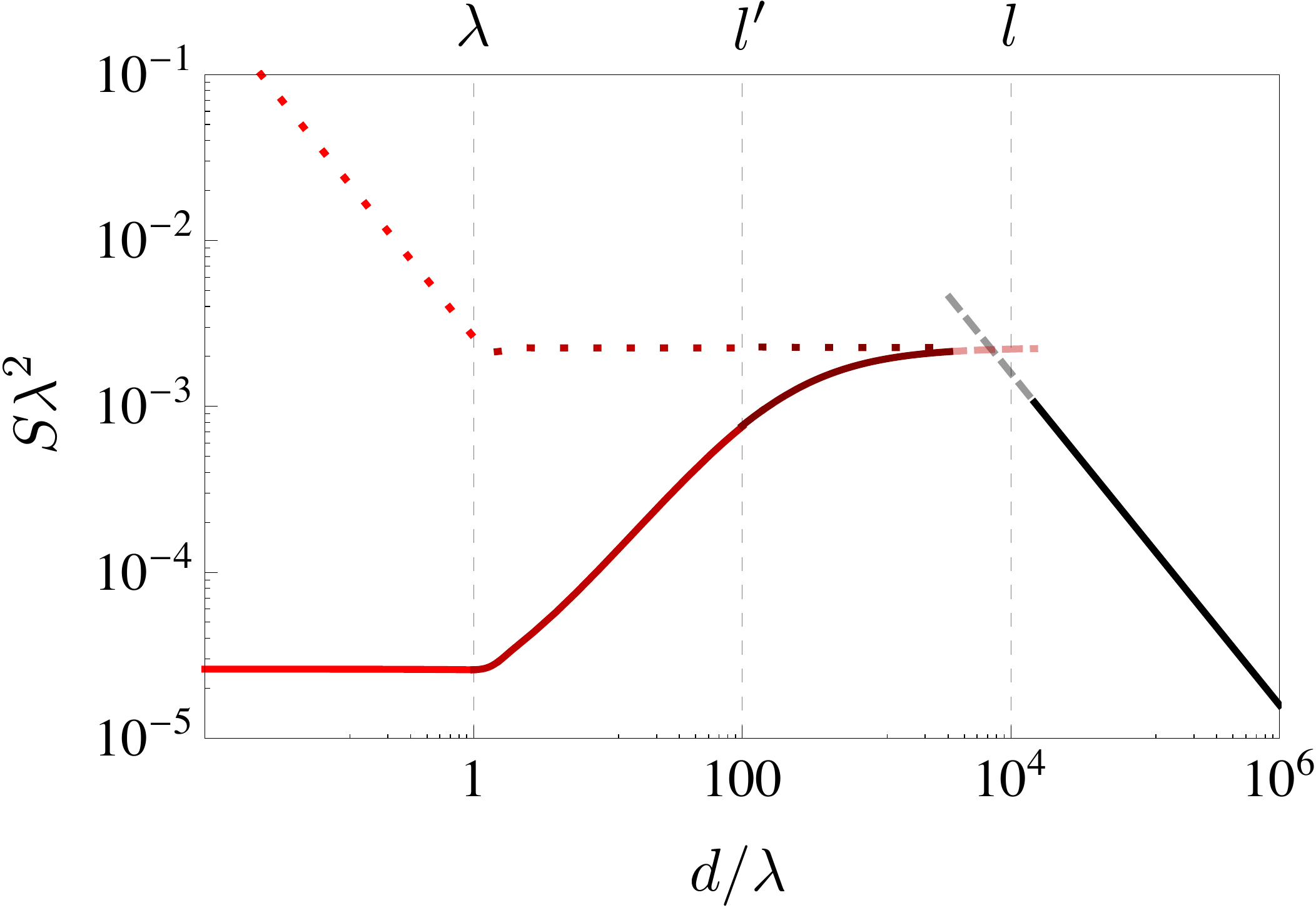}
\caption{Plot of the spin Seebeck coefficient, Eq.~\eqref{SS}, as a function of ferromagnet thickness $d$ for $\lambda=1$~nm, $\xi=10$~nm, $l'=100$~nm, and $l=10$~$\mu$m. We use Eq.~(\ref{SDgL}) (solid curve on the left) and Eq.~(\ref{Sdinfty}) (solid line on the right), which are valid when $d\lesssim l$ and $d\gtrsim l$, respectively, when N$_1$ is a poor spin sink and N$_2$ a perfect spin sink. To account for the classical-to-quantum crossover, we have inserted factor \eqref{QF} in the integrands of Eqs.~\eqref{S0} and \eqref{Sn} [with $\omega\to\omega_n(q)$] and Eq.~\eqref{Sdinfty}. The dotted curve shows the enhanced spin Seebeck coefficient when also N$_1$ is a perfect spin sink (which increases the effective thermal bias between magnons and electrons at the F$\mid${N$_2$} interface).}
\label{SG}
\end{figure}

We summarize the spin Seebeck coefficient dependence on the ferromagnetic layer thickness, when $\lambda\ll l'\ll l$:
\eq{
S(d) \sim\frac{1}{2\pi\lambda^2} \left\{ \begin{array}{cc}
  \alpha \,, & ~~~d\ll\lambda\\
  d/l\,, & ~~~\lambda\ll d\ll l'\\
  l'/l\,, & ~~~l'\ll d\ll l\\
  l'/d\,, & ~~~l\ll d
       \end{array} \right.,
\label{Sd}}
assuming $\lambda\ll\xi$ (or else the magnon transport is exponentially frozen out). The four regimes correspond respectively to the following physical situations: (1) Only the lowest magnon subband is thermally active ($d\ll\lambda$), (2) quasi-3D subband structure is activated, but damping is still dominated by interfacial spin pumping ($\lambda\ll d\ll l'$), (3) bulk damping overtakes spin pumping, but the magnetic film is still thinner than the thermal magnon decay length, such that magnons probe the full film width ($l'\ll d\ll l$), and (4) bulk regime is finally established when the film is thicker than the magnon decay length ($l\ll d$). To illustrate these crossovers, we plot the spin Seebeck coefficient as a function of $d$ in Fig.~\ref{SG}, using lengths characteristic of YIG, which are consistent with the above length-scale hierarchy. Notice that even though $\xi$ determines the magnon energy gap and the associated  Ginzburg-Landau correlation length in the classical theory, it does not govern any prominent crossover in the function $S(d)$.

We conclude that $S(d)$ has nonmonotonic thickness dependence, with the maximum value
\eq{S^{({\rm max})}\sim\frac{\alpha'}{2\pi\lambda^3}\,,\label{Smx}}
attained at $l'\lesssim d\lesssim l$, i.e., below the magnon decay length but above $l'$, such that magnons equilibrate fully to the average phonon temperature $\bar{T}=(T_1+T_2)/2$ ($d\gtrsim l'$) but still remain coherent on the scale of $d$ ($d\lesssim l$). This agrees with the result obtained in Ref.~\onlinecite{xiaoPRB10e}. $S^{({\rm max})}$ is proportional to the spin-mixing conductance [see Eq.~\eqref{ap}] but is independent of the bulk Gilbert damping (as the magnon quasiparticle structure is still well resolved). According to Eq.~\eqref{S},  $S^{({\rm max})}$ determines the largest spin current emitted thermally by a film of magnetic insulator, as a function of $d$, when subjected to a certain temperature \textit{difference} (for example, in a wedged magnetic insulator coated by metallic contacts). If, on the other hand, a well-defined temperature \textit{gradient} is supplied (corresponding, for example, to a certain phonon-dominated heat-flux density), while thickness $d$ is varied, the spin-current density $j_s\propto Sd$ increases with $d$ saturating at $d\gtrsim l$ (the magnon decay length):
\eq{\left.j_s^{({\rm max})}\right|_{{\rm fixed}~\partial_xT}\sim \frac{l'}{2\pi\lambda^2}k_B\partial_x T\,,\label{jm}}
which corresponds to the bulk regime. $j_s^{({\rm max})}$ vanishes when $\alpha'\to0$ (no spin pumping) or $\alpha\to\infty$ (no magnetic dynamics).

It is interesting to ask how the above results would modify if both N$_ 1$ and N$_2$ in our model (see Fig.~\ref{sc}) were perfect spin sinks. For an inversion-symmetric structure (e.g., Pt$\mid$YIG$\mid$Pt), spin currents at the two interfaces must be equal, $j_{s1}=j_{s2}\equiv j_s$. When $d\gg l$, we should recover the bulk limit \eqref{Sdinfty}, since the magnons decay before traversing the full width of the film (and thus the spin current at one interface should not be sensitive to the boundary condition at the other). When $d\ll l$, however, $S_0$ and $S_n$ entering Eq.~\eqref{SDgL} need to be modified. To that end, we notice that the factor $(1+l'/d)^{-1}$ in Eq.~\eqref{S0} reflects the difference between $T_{m,0}$, the effective temperature of the magnons, and $T_2$, the temperature of the electrons in N$_2$:\cite{xiaoPRB10}
\eq{T_{m,0}-T_2=\frac{\alpha \bar{T}+(\alpha'/d)T_2}{\alpha+\alpha'/d}=\frac{T_1-T_2}{2}\frac{1}{1+l'/d}\,.\label{T0}}
Similarly for the $n>0$ subbands, the factor $(1+2l'/d)^{-1}$ in Eq.~\eqref{Sn} stems from the effective magnon-electron temperature difference across the F$\mid$N$_2$ interface of
\eq{T_{m,n}-T_2=\frac{T_1-T_2}{2}\frac{1}{1+2l'/d}\,.\label{Tn}}
When the bulk damping dominates over the interfacial spin pumping, i.e., $l'\ll d$ (while still $d\ll l$), $T_{m,n}\to(T_1+T_2)/2$, while in the opposite limit, i.e., $d\ll l'$, $T_{m,n}\to T_2$. The magnon temperature thus becomes strongly skewed toward the F$\mid${N$_2$} interface for thinner ferromagnetic films, when N$_1$ is a poor spin sink (with electrons and magnons, therefore, being essentially decoupled at the {N$_1$}$\mid$F interface, for our purposes). In the case when both N$_1$ and N$_2$ are perfect spin sinks, on the other hand, the effective magnon-electron temperature difference driving spin current is given simply by $\bar{T}-T_2=(T_1-T_2)/2$ for all subbands (when $d\ll l$). We account for this increased thermal gradient by dropping the factor $(1+l'/d)^{-1}$ on the right-hand side of Eq.~\eqref{S0} and likewise $(1+2l'/d)^{-1}$ in Eq.~\eqref{Sn}. The corresponding enhancement of the spin Seebeck coefficient reverses the trend in Eq.~\eqref{Sd} at $d\lesssim l'$ to give
\eq{
S(d) \sim\frac{1}{2\pi\lambda^2} \left\{ \begin{array}{cc}
  \alpha'/d \,, & ~~~d\ll\lambda\\
  l'/l\,, & ~~~\lambda\ll d\ll l\\
  l'/d\,, & ~~~l\ll d
       \end{array} \right.,
\label{Sds}}
which is now monotonically decreasing with $d$, as plotted by the dotted line in Fig.~\ref{SG}.

\begin{figure}[pt]
\includegraphics[width=\linewidth,clip=]{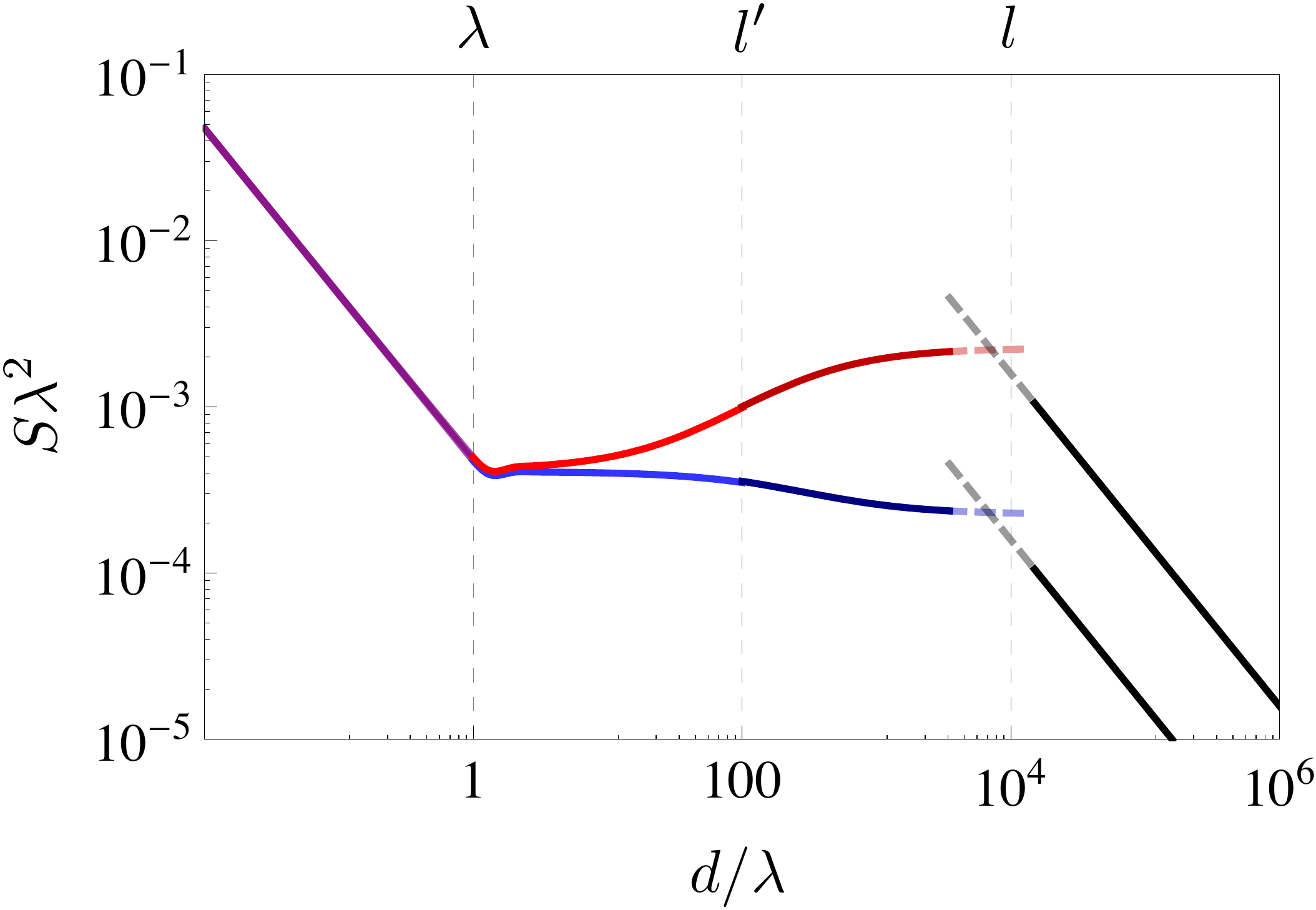}
\caption{Plot of the spin Seebeck coefficient using magnetic length scales as in Fig.~\eqref{SG} but calculated for a non-inversion-symmetric {N$_1$}$\mid$F$\mid${N$_2$} structure with spin-pumping parameters $\alpha_1'=\alpha'/10$, $\alpha_2'=\alpha'$, respectively, at the two interfaces (upper trace) and $\alpha_1'=\alpha'$, $\alpha_2'=\alpha'/10$ (lower trace). Note that the former case is intermediate between the two curves plotted in Fig.~\ref{SG} (where the solid curve corresponds to $\alpha_1'=0$, $\alpha_2'=\alpha'$ and the dotted curve to $\alpha_1'=\alpha_2'=\alpha'$).}
\label{AS}
\end{figure}

When the structure {N$_1$}$\mid$F$\mid${N$_2$} is not mirror symmetric (either because the spin-mixing conductances or the spin-sink characteristics are different), which we characterize by different $\alpha_1'$ and $\alpha_2'$ spin-pumping parameters at the two interfaces, we can repeat the above analysis for $d\ll l$, finding
\begin{align}
T_{m,0}-T_2&=\frac{\alpha \bar{T}+(\alpha_1'/d)T_1+(\alpha_2'/d)T_2}{\alpha+\alpha_1'/d+\alpha_2'/d}\nonumber\\
&=\frac{T_1-T_2}{2}\frac{1+2l_1'/d}{1+(l_1'+l'_2)/d}
\end{align}
and
\eq{T_{m,n}-T_2=\frac{T_1-T_2}{2}\frac{1+4l_1'/d}{1+2(l_1'+l_2')/d}\,,}
where we defined the spin-pumping lengths $l_i'\equiv\alpha_i'/\alpha$ associated with the left, $i=1$, and right, $i=2$, interfaces. We thus generalize the spin Seebeck contributions (at the F$\mid$N$_2$ interface) from different magnon subbands to
\eq{S_0\approx\frac{\alpha_2'}{d}\frac{1+2l_1'/d}{1+(l_1'+l_2')/d}\int_{-\infty}^\infty\frac{d^2\textbf{q}}{(2\pi)^2}F[\beta\hbar\omega_0(q)]}
and
\eq{S_{n>0}\approx\frac{2\alpha_2'}{d}\frac{1+4l_1'/d}{1+2(l_1'+l_2')/d}\int_{-\infty}^\infty\frac{d^2\textbf{q}}{(2\pi)^2}F[\beta\hbar\omega_n(q)]}
in lieu of Eqs.~\eqref{S0} and \eqref{Sn}, respectively. The asymptotic and crossover trends are now given by (assuming $\lambda\ll l_1',l_2'\ll l$):
\eq{
S(d) \sim\frac{1}{2\pi\lambda^2} \left\{ \begin{array}{cc}
  2/({\alpha'_1}^{-1}+{\alpha'_2}^{-1})d \,, & ~~~d\ll\lambda\\
  2/({l'_1}^{-1}+{l'_2}^{-1})l\,, & ~~~\lambda\ll d\ll l_1'\\
  l_2'/l\,, & ~~~l_1',l_2'\ll d\ll l\\
  l_2'/d\,, & ~~~l\ll d
       \end{array} \right.,
}
which coincides with Eq.~\eqref{Sds} when $\alpha'_1=\alpha'_2$ but has an additional shoulder-like feature at $d\sim (l_1'+l_2')/2$ . This feature makes $S(d)$ nonmonotonic when $\alpha'_1<\alpha'_2$. In Fig.~\ref{AS}, we plot the spin Seebeck coefficient for two asymmetric cases: (1) $\alpha_1'=\alpha'/10$, $\alpha_2'=\alpha'$ and (2) $\alpha_1'=\alpha'$, $\alpha_2'=\alpha'/10$ (physically corresponding to spin currents on two sides of a non-inversion-symmetric {N$_1$}$\mid$F$\mid${N$_2$} structure, such as, Pd$\mid$YIG$\mid$Pt).

\section{Discussion}
\label{DS}

Our theory provides a minimalistic application of the LLG phenomenology to the problem of the spin Seebeck effect, yet disregards magnon-magnon interactions. These become important at high temperatures, especially approaching the Curie temperature. Magnon-phonon interactions are included only insofar as a contribution to the total Gilbert damping. Elastic magnon scattering on impurities, which would manifest as an inhomogeneous broadening of ferromagnetic-resonance linewidth, may be an important impediment to the thermally-induced spin currents in disordered films. The bulk limit of spin current, Eq.~\eqref{jm}, is reduced by disorder, as well as the magnon decay length describing the crossover to the bulk regime. When the magnon mean free path $l^\ast$ is shorter than our Gilbert damping decay length $l$, in particular, we expect the effective decay length to be $l_{\rm eff}\sim\sqrt{ll^\ast}$ (the spin diffusion length) and $j_s^{({\rm max})}$ in Eq.~\eqref{jm} to be reduced by a factor of $l/l_{\rm eff}\sim\sqrt{l/l^\ast}$ (assuming that $l'\ll l_{\rm eff}$, such that our length-scale hierarchy is unchanged). The Seebeck coefficient behavior \eqref{Sd} [as well as Eq.~\eqref{Smx}], however, remain essentially intact up to the thickness $d\sim l_{\rm eff}$.

Finally, we want to comment on a possibility of nonlocal magnetic relaxation. In this paper, we have assumed that Gilbert damping is a local and isotropic tensor. The locality would be a reasonable approximation if the damping bottleneck was due to some local dynamic defects. In the case of YIG,\cite{cherepanovPRP93} which is known for its highly coherent elastic properties, nonlocality of the bulk magnetic relaxation could significantly modify our findings. First of all, this could introduce new phonon-dependent length scales into the problem, which would show in the $S(d)$ dependence. Standard long-wavelength ferromagnetic resonance on thick films would reveal damping $\alpha$ that could be very different from that of short-wavelength thermal magnons relevant here. An effective damping parameter $\tilde{\alpha}$ of thermal magnons (which may itself be thickness dependent) would result in the bulk crossover thickness of $\tilde{l}\sim\lambda/\tilde{\alpha}\neq l$. When $d\lesssim\tilde{l}$, we may still invoke the first three regimes of our findings, Eq.~\eqref{Sd} (with $\alpha$ and $l$ corresponding to the thermal-magnon inverse quality factor and decay length, respectively, due to magnon-phonon scattering), which should, furthermore, be indifferent to the fact that the local temperature, Eq.~\eqref{Tx}, is not well defined for highly-coherent phonons. The reason for this is that, in these regimes, when $d$ is below the magnon decay length $\tilde{l}$, only the average phonon temperature $\bar{T}$ is relevant for our theory. Our bulk regime, $S(d)\propto d^{-1}$, when $d$ is larger than the magnon decay length, would, however, have to be considerably revised in the case of nonlocal magnetic relaxation, calling for a rigorous quantum-kinetic theory.\cite{cherepanovPRP93}

In summary, we have developed a minimal Landau-Lifshitz theory of the longitudinal spin Seebeck effect, which calls for its systematic experimental study for the temperature and film-thickness dependence, which, in turn, may necessitate a more systematic microscopic quantum-kinetic theory.

This work was partially supported by a Grant No. 228481 from the Simons Foundation, the NSF under Grant No. DMR-0840965, and FAME (an SRC STARnet center sponsored by MARCO and DARPA). The authors acknowledge stimulating discussions with G. E. W. Bauer and J. Xiao.

\end{document}